\documentstyle [12pt] {article} 
\newcommand{\sect}[1]{ \section{#1} } 
 
\newcommand{\req}[1]{(\ref{#1})}

\newcommand{\ve}{\left( \begin{array}{r}} 
\newcommand{\ev}{\end{array} \right)} 
\newcommand{\ar}{\left( \begin{array}{ll}} 
\newcommand{\ra}{\end{array} \right)} 
\newcommand{\arr}{\left( \begin{array}{rrrr}} 
\newcommand{\arrr}{\left( \begin{array}{rrrrrr}} 

\newcommand{\eqr}{\begin{eqnarray}}
\newcommand{\rqe}{\end{eqnarray}} 
\newcommand{\eq}{\begin{equation}} 
\newcommand{\qe}{\end{equation}}

\newcommand{\half}{\frac{1}{2}}

\newcommand{\ket}[1]{\mbox{$ \mid #1 >$}}

\newcommand{\eps}{\epsilon}

\newcommand{\vol}[1]{d{#1}\wedge d\bar{{#1}}}

\begin{document}

\begin{flushright}
HUTP-97/A091\\
hep-th/9711083
\end{flushright}

\vspace{2cm}

\begin{center}

{\bf\Large
Extended Objects in MQCD
} 

\vspace{1.5cm}

Ansar Fayyazuddin and Micha\l\ Spali\'nski
\footnote{On leave from the
Institute of Theoretical Physics, Warsaw University, Poland.}\\
\vspace{0.2cm}

{\em
Lyman Laboratory of Physics\\
Harvard University\\
Cambridge, MA 02138, U.S.A.
}
\end{center}

\vspace{1.4cm}

\begin{abstract} 

$N=1$ supersymmetric QCD is considered using the recently proposed picture
of it as the world volume theory of a single M-theory fivebrane with two of
its dimensions wrapped on a Riemann surface. The conditions under which a
second M-theory brane can be introduced preserving some supersymmetry are
analysed. Such configurations represent BPS saturated extended objects in
the quantum field theory on the brane worldvolume.  Formulae for the
tension of these extended objects are derived. An explicit intersecting
fivebrane configuration is found which is interpreted as a BPS domain wall
in $4$-dimensional MQCD.

\end{abstract}

\thispagestyle{empty}

\newpage

\setcounter{page}{1}

\sect{Introduction} 

One area of research where non-perturbative string and M-theory
ideas\footnote{For recent reviews see \cite{klemm} and 
\cite{mtheory} .} have found fruitful application
is the study of non-perturbative aspects of gauge theories. It has given a
new and interesting perspective on non-perturbative aspects of quantum
field theory and the relation between the classical and the quantum. The
string/M-theory viewpoint has provided a natural physical interpretation
for the Seiberg-Witten solution \cite{sw1,sw2} of $N=2$ supersymmetric
Yang-Mills theory. 

Recently BPS saturated extended objects in $N=1$ 
supersymmetric quantum field theory have attracted much 
attention \cite{dva,kov,chi}. This article investigates the M-theory
interpretation of these objects as BPS stable 
configurations of intersecting M-branes. 

In the M-theory setting one can view $4$-dimensional quantum field theory
as the worldvolume theory of an M-theory fivebrane wrapped on a Riemann
surface $\Sigma$ \cite{klmvw,wit1}. Extended objects in the quantum
field 
theory constructed in this fashion can naturally be interpreted as other
branes intersecting this ``vacuum'' fivebrane in various ways. Depending on
how the worldvolumes intersect the corresponding state in the field theory
will have a very different appearance. Supergravity and worldvolume
effective field theory considerations suggest \cite{gau,tse,tow,stro} that
a pair of fivebranes can 
intersect along a space of either $3$ or $1$ dimensions, while a fivebrane 
and a twobrane can intersect along $1$ dimension. In the
present analysis these results are not assumed. In fact, to study extended
objects in the effective quantum field theory on the worldvolume it seems
most natural to classify the intersections by the number of dimensions of
the intersecting fivebrane that coincide with the non-compact $3$
dimensional (space) part of the ``vacuum'' fivebrane. These intersections
can be interpreted as membranes ($d=2$), strings ($d=1$) or particles
($d=0$) in the $4$-dimensional quantum field theory on the ``vacuum''
fivebrane worldvolume. This paper is devoted to the study of BPS stable
configurations of this kind.

It is shown that only two types of intersections can lead to solutions of
the BPS conditions: the fivebrane-fivebrane intersection along a $2$-space
in $3+1$ dimensional spacetime, which has a natural domain wall
interpretation, and the fivebrane-fivebrane intersection along a $1$-space,
which would be 
interpreted as a BPS saturated string. The explicit membrane solution
found here 
is in fact a $3$-dimensional intersection (in agreement with expectations
based on supergravity).  However the string solution seems singular, as it
requires that the volume (hence also the tension) of this object should
vanish in the BPS limit.

Section $2$ reviews the M-theory fivebrane configurations leading to the
gauge theories of interest, following the original presentation of Ooguri
et al. \cite{hori} and Witten \cite{wit2} (as well as the papers
\cite{biksy,witek,nam,raman,ahn1,ahn2,ahn3,ahn4,boer,oog} which generalise
and explore the original construction). Various consistency checks like
decoupling and Seiberg duality are discussed.

Section $3$ is devoted to counting the supersymmetries of the fivebrane
configuration using the methods of \cite{bbs}. Section $4$ considers the
various fivebrane-fivebrane intersections, and section $5$ discusses an
explicit example of a BPS domain wall in SU(n) MQCD. Finally section $6$
summarises the results for fivebrane-twobrane intersections.

\sect{M-theory Fivebranes with Four Supersymmetries}

Four dimensional $N=1$ supersymmetric $SU(n)$ Yang-Mills theory can be
regarded as the world-volume theory of parallel fourbranes in type IIA
string theory with finite extent in one direction, $x^6$.  The ends of the
fourbranes, along the $x^6$ axis, lie on fivebranes.  The remaining
coordinates of the world-volume theory of the fourbranes are $x^0, x^1,
x^2, x^3$.  

In the $N=2$ case the fivebranes on which the fourbranes end have
world-volumes extending 
in the $x^0,x^1,x^2,x^3,x^4,x^5$ coordinates. This configuration preserves
a quarter of the 32 real supersymmetries of M-theory, that is eight. To
have $N=1$ supersymmetry in four dimensions one has to break a half of
these. This can be achieved by rotating one of the fivebranes so that they
are not parallel anymore\cite{eli1,eli2,bsty}. Thus one fivebrane extends over
$x^0,x^1,x^2,x^3,x^4,x^5$ while the other extends over
$x^0,x^1,x^2,x^3,x^7,x^8$. 

To understand the physical limit under consideration here it is useful to
first recall the $N=2$ situation. The field theory on the fourbrane worldvolume
is a supersymmetric 
Yang-Mills theory in a limit where gravity decouples. The idea of
decoupling gravity in order to understand non-perturbative aspects of
field 
theory from the string point of view was first made explicit in
\cite{kklmv}. It was subsequently understood 
that the field theory limit only depends on the local structure of the
compactification and can be explored by ``engineering'' the singularity
structure appropriately \cite{klmvw,kkv,kmv}. The analog of this in
the brane language was developed in a number of papers
\cite{hw,wit1} and the relationship between the  
alternative approaches has been discussed in \cite{ov} and very recently in
\cite{val}. Gravity is decoupled by taking the string coupling  $\lambda$
to be small. Since the Yang-Mills coupling $g$ on the fourbrane worldvolume
theory satisfies
\eq
{1\over g^2} = {L\over \lambda\l_s}
\qe
(here $l_s$ is the string scale), one has to keep this finite to get a
nontrivial limit. Thus, as the string coupling is taken to be small,
keeping $g$ fixed entails keeping the ratio $L/(\lambda l_s)$ fixed. 
Since $R=\lambda l_s$, this is the same as fixing $L/R$. 
This is intuitively sound, as it also
means that the Kaluza-Klein excitations decouple in the limit. 

Using an M-theory fivebrane configuration instead of
the flat branes in type IIA string theory means working in the strongly
coupled regime. In \cite{wit1} it is argued that the physical picture in
the dimensions $(0,1,2,3)$ persists also when 
this {\em opposite} limit is taken, that is, one takes the string coupling
$\lambda$ to be large keeping the Yang-Mills coupling finite,
i.e. $R$ and $L$ are taken large keeping their ratio fixed. This 
does not change the physics, because the theory on the Coulomb branch
depends on $R$ and $L$ only via their ratio (see also \cite{oog}). 

A similar picture is advocated in \cite{hori,wit2,biksy} for the case of $N=1$,
although it is also pointed out \cite{wit2} that in this case gravity does
not entirely 
decouple, but instead leads to a one-parameter generalisation of standard
supersymmetric Yang-Mills theory. As in the $N=2$ case, 
since the fourbrane is just an M-theory fivebrane wrapped around $x^{10}$,
one can ask what the M-theory fivebrane configuration would have to be
to lead to the type IIA configuration of fourbranes and fivebranes
described above in the weakly coupled limit. 
The answer is that the fivebrane fills the space parameterised by 
$x^0,x^1,x^2,x^3$ and its two remaining dimensions are wrapped on a Riemann
surface $\Sigma$ which is
holomorphically embedded in $R^6$ (the $x^4,x^5,x^6,x^7,x^8,x^{10}$
coordinates). Adopting a complex structure where $s=x^6+i x^{10}$, $v=x^4+i
x^5$, $w=x^7+i x^8$ are holomorphic, the Riemann surface $\Sigma$ will be
given by two holomorphic relations in the variables $s, v, w$. Since $x^{10}$
is really compactified on a circle, instead of $s$ one uses
$t=exp(-s/R)$, $R$ being the radius of the compact $x^{10}$. The radius $R$
will be set to unity in what follows, except in section \req{example}.

One way of
finding the curve $\Sigma$ is to emulate the effect of rotating one of the
fivebranes in the M-theory limit. This can be done at points in $N=2$ moduli
space where the curve degenerates to genus zero\cite{hori,wit2,biksy}. At
such points the curve 
can be parameterized rationally, which means that it can be specified by
prescribing $t,v,w$ as rational functions of a coordinate $\lambda$ on the
complex sphere. To determine these functions one has to examine the
behaviour of $t,v,w$ to see what kind of singularities are allowed. 

In the $N=2$ case $v$ should have two poles corresponding to the two
fivebranes, both at $w=0$. After the rotation, there should be a pole in
$v$ at $w=0$ (this corresponds to the ``left'' fivebrane), and one in $w$
at $v=0$ (the other fivebrane). This implies the relation
\eq
wv=\zeta
\qe
for some $\zeta$. To specify the curve completely one more relation is
needed. 

To discuss all the cases in a self-contained way, consider the
$N=2$ curve given by
\eq
y^2 = A(x)^2 - B(x).
\qe
This form is common for all the non-exceptional classical groups (see
e.g. \cite{hok}):
\eqr
SU(r+1): A(x) &=& \prod_{k=1}^{r+1} (x-\bar{a}_k), \nonumber \\
B(x) &=&\Lambda^{2r+2-N_f}\prod_{j=1}^{N_f} (x+\mu_j) ; \\
SO(2r+1): A(x) &=& \prod_{k=1}^{r} (x^2-\bar{a}^2_k), \nonumber \\
B(x) &=&\Lambda^{4r-2N_f-2}x^2 \prod_{j=1}^{N_f} (x^2-\mu_j^2) ; \\
Sp(2r): A(x) &=& x^2 \prod_{k=1}^{r} (x^2-\bar{a}^2_k) + A_0 , \nonumber \\
B(x) &=&\Lambda^{4r-2N_f-2}x^2 \prod_{j=1}^{N_f} (x^2-\mu_j^2) ; \\
SO(2r): A(x) &=& \prod_{k=1}^{r} (x^2-\bar{a}^2_k), \nonumber \\
B(x) &=&\Lambda^{4r-2N_f-2} x^4 \prod_{j=1}^{N_f} (x^2-\mu_j^2) .
\rqe
Here $\mu_i$ are the flavour masses, $\Lambda$ is the renormalisation group
invariant scale, $\bar{a}_k$ are the branch points, and
\eq
A_0 = \Lambda^{2r-N_f+2}\prod_{j=1}^{N_f} \mu_j. 
\qe
Setting $y=t+A(x)$, $x=v$ one finds
\eq
t^2 + 2 t A(x) + B(x) = 0.
\qe 
By rescaling $x$ and $B$ suitably (and identifying $x\equiv v$) one can assume
\eqr
A(v) &=& v^a + \dots \nonumber \\
B(v) &=& \mu v^b + \dots
\rqe

The idea is to consider a degenerate limit, where the Riemann surface
becomes a sphere, and read off the asymptotic behaviour of $t$. 
In the limit $v\rightarrow\infty$ the curve becomes:
\eq
t^2 + tv^{a} + \mu v^b = 0 .
\qe
For the $N=2$ curves $b\leq 2a$, which is assumed throughout. One can  
distinguish different regions of interest for the parameters
$b,a$.  If $b<2a$, then:
\eqr
\label{asymp1}
t & \rightarrow & -\mu v^{b-a} , \nonumber \\
\mbox{or  }
t & \rightarrow & -v^a \rightarrow \infty .
\rqe
The first root has the following asymptotic behaviour as $v\rightarrow\infty$:
\eqr
\label{asymp2}
b<a \quad & = & t\rightarrow -\mu v^{b-a}\rightarrow 0 , \nonumber\\
b=a \quad & : & t\rightarrow  -\mu v^{b-a}\rightarrow -\mu , \nonumber\\
b>a \quad & : & t\rightarrow  -\mu v^{b-a}\rightarrow\infty .
\rqe
When $b=2a$:
\eq
\label{asymp3}
t\rightarrow \frac{1}{2}v^a(-1\pm\sqrt{1-4\mu})\rightarrow
\infty . 
\qe
Next the behaviour in the vicinity of the zeros of $B$ (denoted by $\mu_i$)
is needed. Suppose 
\eq B=\mu\prod_{i=1}^p (v-\mu_i)^{n_i}, \qquad
\sum_{i=1}^p n_i = b.  
\qe
Expanding $A$ around  $\mu_i$ gives
\eq
A(v) = C_i (v-\mu_i)^{m_i} + \dots
\qe
for some constant $C_i$ (note that $m_i$ are allowed to vanish).
Now concentrate on a point in moduli space where the Riemann surface
degenerates to a sphere; there the surface can be parameterized by
a single 
parameter $\lambda$ \cite{hori,wit2}.  Let
\eq 
v = \lambda + \frac{d}{\lambda},
\qe
so that $v\rightarrow\infty$ as $\lambda\rightarrow\infty$ and
$\lambda\rightarrow 0$.  These two infinities correspond to the 
two five-branes between which the fourbranes, in the weak coupling
type IIA picture, are suspended.  
Since the asymptotic behaviour of $t$ as $v\rightarrow\infty$ 
is determined by a quadratic equation, it has two different behaviours
corresponding to the two roots of the equation as given in 
(\ref{asymp1},\ref{asymp2},\ref{asymp3}). One may associate these two roots
with either $\lambda\rightarrow\infty$ or 
$\lambda\rightarrow 0$.  Let us denote by 
(A) the case when the first root given in \req{asymp1} is associated with 
$\lambda\rightarrow 0$ and the second one with
$\lambda\rightarrow\infty$. Then 
\eqr
\lambda\rightarrow 0\quad & : & t\rightarrow
-\mu \left(\frac{\lambda}{d}\right)^{b-a} , \nonumber\\
\lambda\rightarrow\infty\quad & : & t\rightarrow -\lambda^{a} .
\rqe
The function $t(\lambda )$ has a pole of order $a$ at
$\lambda\rightarrow\infty$ and a pole of order $b-a$ at $\lambda = 0$
if $b<a$ or a zero of order $a-b$ at $\lambda =0$ if $b>a$.    
Since $\lambda$ parameterizes a sphere, $t$ should have an equal number
of poles and zeroes.  Thus $t$ must have a further set of zeroes whose
total order should be $b$.  These occur whenever the polynomial $B$
vanishes, i.e. whenever
\eq
v = \lambda + \frac{d}{\lambda} = \mu_i .
\qe  
Denoting the two roots of these equations as $\lambda_{\pm i},
\lambda_{+ i}\lambda_{- i}=d $,
one can write $t(\lambda )$ uniquely as:
\eq
t = -\prod_{j=1}^{k}(\lambda -\lambda_{+ j})^{n_{j} - m_{j}}
(\lambda - \lambda_{- j})^{m_{j}}\lambda^{a-b},
\qe
where the $m_j, n_j$ are defined above and $k$ is the number of distinct
zeroes of the polynomial $B$. One also obtains the equation:
\eq
\label{cons}
\mu = (-1)^{b}d^{a-b}\prod_{j}^{k}\lambda_{- j}^{2m_{j}-n_{j}}d^{n_{j}-m_{j}}.
\qe

Now consider decoupling flavors with mass $\mu_j$
by taking $\mu_{j}\rightarrow\infty$ while
keeping $\mu\mu_{j}^{n_{j}}$ fixed in the usual way.  In this limit 
$\lambda_{- j}\rightarrow -\mu_{j}$ and one can only keep the equations
consistent with the case of a smaller number of flavors if
$m_{j}=0$.  
Thus decoupling requires that the theory be at a point in moduli space at
which $m_j =0$, so finally one finds:
\eqr
t(\lambda)&=& -\prod_{j=1}^{k}(\lambda - \lambda_{+ j})^{n_j}\lambda^{a-b}
\nonumber \\
\mu &=& (-1)^{b}d^{a-b}\prod_{j}^{k}\lambda_{+ j}^{n_j}.
\rqe

\noindent Rotating one of the fivebranes
to break the supersymmetry by a further half corresponds to
keeping only one of the infinities in $v$, and letting the other
one correspond to an infinity in $w$.  Thus one can take 
\eqr
v & = &\lambda , \nonumber\\
w & = &\frac{\zeta}{\lambda},
\rqe
which yields the $N=1$ curve:
\eqr
t &=& -\prod_{j=1}^{k}(v - v_{j})^{n_j} v^{a-b} ,
\nonumber \\
vw &=& \zeta .
\rqe

Let us now consider the other possible case, denoted by 
(B), in which one associates the second root given in \req{asymp1} to 
$\lambda\rightarrow 0$ and the first root to $\lambda\rightarrow\infty$.
In that case 
\eqr
\lambda\rightarrow 0\quad & : & t\rightarrow
\left(\frac{d}{\lambda}\right)^{a} , \nonumber\\
\lambda\rightarrow\infty\quad & : & t\rightarrow -\mu\lambda^{b-a} .
\rqe
The same argument as for the case (A) leads to the $N=1$ curve:
\eqr
t &=& -\prod_{j=1}^{k}(v - {\tilde v}_{j})^{n_j} v^{-a} ,
\nonumber \\
vw &=&\zeta.
\rqe
Thus one arrives at two different curves describing the same physical
situation, related by the discrete transformation: $a\rightarrow a-b, 
b\rightarrow b$. This is exactly Seiberg's non-abelian duality.

\sect{The MQCD vacuum}

MQCD is the quantum field theory living on the $(0,1,2,3)$ part of a
fivebrane with world 
volume $(0,1,2,3)\times\Sigma$ where $\Sigma$ is a Riemann surface
holomorphically embedded in $Y \equiv R^5\times S^1$ as
described in the preceding section.

The following sections consider the possibility of having BPS saturated
extended objects in the world volume theory of the fivebrane.  Three types
of objects are studied, which intersect 
the $(0,1,2,3)$ part of the 
fivebrane world volume at $0,1$ or $2$ dimensional 
manifolds.  If these objects exist, they appear as BPS saturated particles,
strings, and membranes in MQCD. To establish whether they do exist, one has
to look for brane configurations which preserve some supersymmetry.
The first step is to verify the number of 
supersymmetries preserved 
by a single fivebrane with world volume $(0,1,2,3)\times\Sigma$, that is, 
the number of supersymmetries preserved by the vacuum in MQCD. This
calculation is necessary, since it is important for the following sections
to have an explicit form for the preserved supersymmetry
transformations, that is, for the spinorial parameters of these
supersymmetries. 

The conditions for preservation of supersymmetry can be found following
\cite{bbs} (see also \cite{cve,fs}). The number of supersymmetries
preserved by 
a p-brane configuration, whose embedding in $R^{9,1}\times S^1$ is
described by the maps $X^{M}$, is given 
by the number of spinors $\chi$ which satisfy the equation 
\eq
\label{memb}
\chi = \frac{1}{p!}\eps^{\alpha_1 \dots \alpha_p}
\Gamma_{M_1\dots 
M_p} \partial_{\alpha_1} X^{M_1} \dots \partial_{\alpha_p} X^{M_p}
\chi .
\qe
Here \footnote{The antisymmetrization in \req{muga} includes a factor of
$1/p!$, and the epsilon symbol in \req{memb} is a tensor, not a tensor
density.}  
\eq
\label{muga}
\Gamma_{M_1\dots M_p} = \Gamma_{[M_1}\dots\Gamma_{M_p]}
\qe
and $\Gamma_M$ are $11$-dimensional gamma matrices.
 
For a fivebrane with world-volume filling $x^0,....,x^3,\Sigma$ the
supersymmetry condition (in the static gauge) reduces to: 
\eq 
\label{susyv}
\chi =
\frac{1}{2}\eps^{\alpha \beta} \Gamma_{0}\dots \Gamma_{3} \Gamma_{ij}
\partial_{\alpha} X^i \partial_{\beta} X^j \chi , 
\qe 
where $i,j$ label the
embedding coordinates $(X^4,X^5,X^6,X^7,X^8,X^{10})$.  Using complex
coordinates\footnote{Also $T=\exp(-S/R)$ will be used.}
\eqr 
S&=&X^6+iX^{10}, \nonumber \\ 
V&=&X^4+iX^5 , \nonumber \\ 
W&=&X^7+iX^8 , \\ 
\rqe
which will be denoted by $X^m$ (with $X^{\bar{m}}$ as their complex
conjugates), the condition \req{susyv} reads
\eqr 
\chi &=& \eps^{\alpha \beta} \Gamma_{0}\Gamma_{1}\Gamma_{2}\Gamma_{3}
(\Gamma_{mn} \partial_{\alpha} X^m \partial_{\beta} X^n + \Gamma_{m{\bar
n}} \partial_{\alpha} X^m \partial_{\beta} X^{\bar n} + 
\Gamma_{{\bar m}n} \partial_{\alpha} X^{\bar m} \partial_{\beta} X^{n} , 
\nonumber \\
&+&\Gamma_{{\bar m}{\bar n}} \partial_{\alpha} X^{\bar m} 
\partial_{\beta} X^{\bar n}) \chi .
\rqe 
As the fivebrane is wrapped around the holomorphic curve $\Sigma$, only
the term with one holomorphic and one anti-holomorphic index is
non-vanishing: 
\eq 
\chi = \eps^{\alpha\beta} \Gamma_{0}\dots
\Gamma_{3} (\Gamma_{m\bar{n}} \partial_{\alpha} X^m \partial_{\beta}
X^{\bar{n}}+ \Gamma_{{\bar m}n} \partial_{\alpha} X^{\bar m} 
\partial_{\beta} X^{n}) \chi .  
\qe
This equation can be rewritten as:
\eq 
\sqrt{h}d\sigma^{1}\wedge d\sigma^{2}\chi = 
\Gamma_{0}\Gamma_{1}\Gamma_{2}
\Gamma_{3}\Gamma_{m\bar{n}} (\partial_{1} X^m \partial_{2}
X^{\bar{n}}-\partial_{2} X^m \partial_{1}
X^{\bar{n}})d\sigma^{1}\wedge d\sigma^{2} \chi.  
\qe
Here $\sigma^{1,2}$ parameterize the two dimensional surface $\Sigma$ and
$h$ is the determinant of the induced metric on $\Sigma$.
The equations describing the embedding of $\Sigma$ 
in $Y$ 
can be written as:
\eqr
f_1(v,t) & = & 0 , \nonumber\\
f_2(v,w) & = & 0 
\rqe
(the explicit form of the functions $f_1$, $f_2$ was given in section $2$).
From these two equations one can calculate
\eq
h = -\frac{1}{4}(\frac{1}{|T|^2} +|\frac{\partial_{T}f_1}{\partial_{V}f_1}|^2+
|\frac{\partial_{T}f_1}{\partial_{V}f_1}
\frac{\partial_{V}f_2}{\partial_{W}f_2}|^2)^2(\partial_{1}T\partial_{2}
{\bar T}-\partial_{2}T\partial_{1}{\bar T})^2
\qe
and the equation for supersymmetry preservation becomes:
\eqr
&&i\Gamma_{0}\Gamma_{1}\Gamma_{2}\Gamma_{3}(\Gamma_{\bar{T}T}-
\Gamma_{\bar{T}V}\frac{\partial_{T}f_1}{\partial_{V}f_1}+  
\Gamma_{\bar{T}W}(\frac{\partial_{V}f_2}{\partial_{W}f_2}
\frac{\partial_{T}f_1}{\partial_{V}f_1}) - 
\Gamma_{\bar{V}T}(\frac{\partial_{T}f_1}{\partial_{V}f_1})^{\star} , 
\nonumber \\
&+&\Gamma_{\bar{V}V}|\frac{\partial_{T}f_1}{\partial_{V}f_1}|^2 
-\Gamma_{\bar{V}W}|\frac{\partial_{T}f_1}{\partial_{V}f_1}|^2
\frac{\partial_{V}f_2}{\partial_{W}f_2}
+ \Gamma_{\bar{W}T}(\frac{\partial_{T}f_1}{\partial_{V}f_1}
\frac{\partial_{V}f_2}{\partial_{W}f_2})^{\star} ,
\nonumber \\
&-&
\Gamma_{\bar{W}V}|\frac{\partial_{T}f_1}{\partial_{V}f_1}|^2
(\frac{\partial_{V}f_2}{\partial_{W}f_2})^{\star}
+ \Gamma_{\bar{W}W}|\frac{\partial_{T}f_1}{\partial_{V}f_1}
\frac{\partial_{V}f_2}{\partial_{W}f_2}|^2)\chi , \nonumber \\
&=&
\frac{1}{2}(\frac{1}{|T|^2} +|\frac{\partial_{T}f_1}{\partial_{V}f_1}|^2+
|\frac{\partial_{T}f_1}{\partial_{V}f_1}
\frac{\partial_{V}f_2}{\partial_{W}f_2}|^2)(\partial_{1}T\partial_{2}
{\bar T}-\partial_{2}T\partial_{1}{\bar T})\chi .
\rqe
%Since the curve $\Sigma$ is given in terms of two holomorphic constraints ,
%one can relate the various derivatives appearing here. 
This implies the conditions 
\eqr 
\label{sucon1}
i\Gamma_{0}\Gamma_{1}\Gamma_{2}\Gamma_{3} \Gamma_{\bar{T}T} \chi &=& \half
|{1\over T}|^2 \chi , \nonumber \\ 
i\Gamma_{0}\Gamma_{1}\Gamma_{2}\Gamma_{3} \Gamma_{\bar{V}V} \chi &=&
\half \chi  , \nonumber \\ 
i\Gamma_{0}\Gamma_{1}\Gamma_{2}\Gamma_{3} \Gamma_{\bar{W}W}
\chi &=& \half \chi \nonumber \\ 
\rqe 
and 
\eqr
\label{sucon2}
\Gamma_{\bar{T}V}\chi&=&\Gamma_{\bar{T}W}\chi=\Gamma_{\bar{V}W}\chi= 0 , 
\nonumber \\
\Gamma_{\bar{V}T}\chi&=&\Gamma_{\bar{W}T}\chi=\Gamma_{\bar{W}V}\chi= 0 .
\rqe

These conditions admit $4$ real solutions, so $4$ supersymmetries are
preserved, which corresponds to $N=1$ supersymmetry in $4$ dimensions. To
see this it is convenient to regard the gamma matrices for the six
dimensions $4,5,6,7,8,10$ in the present basis
as fermionic creation and annihilation operators, since they satisfy the
anticommutation relations
\eq
\{\Gamma_m, {\Gamma_n}^\dagger\} = g_{m\bar{n}} ,
\qe 
where $g$ is the (flat) metric on $Y$. Thus $\Gamma_m$ is a fermionic
creation operator, while 
its conjugate is a fermionic annihilation operator.

The spinor $\chi$ can
be viewed as a Fock state in the space generated by the operators $\Gamma_m$
acting on a Fock vacuum. It is convenient to denote a vector in the Fock
space by
$\alpha\otimes\ket{n_S,n_V,n_W}$, where the $n_i$ are the fermionic Fock
space occupation numbers (i.e. $0$ or $1$) and $\alpha$ is a
$4$-dimensional Dirac spinor. Thus there are 
$8*2^3=64$  states. The conditions
\req{sucon2} imply that
\eq
\chi = \alpha\otimes \ket{000} + \beta\otimes \ket{111} ,
\qe
where $\alpha,\beta$ are $4$-dimensional (Dirac) spinors. There are
$16$ such states. 

The conditions \req{sucon1} lead to
\eqr
i\Gamma_0\Gamma_1\Gamma_2 \Gamma_3 \alpha_1 &=& \alpha_1 , \nonumber \\
i\Gamma_0\Gamma_1\Gamma_2 \Gamma_3 \alpha_2 &=& - \alpha_2  ,
\rqe
i.e. they determine the $4$-dimensional chirality of the spinors
$\alpha_1,\alpha_2$, leaving $8$ solutions. Finally the eleven-dimensional
Majorana condition must be 
imposed. Recall that with the conventions of \cite{fs} $\Gamma_M$ are
real for $M=2,5,7,8,10$, purely imaginary for $M=0,1,3,4,6,9$, symmetric
for $M=0,2,5,7,8,10$ and antisymmetric for $M=1,3,4,6,9$. The charge
conjugation matrix is given by 
\eq 
C=B\Gamma_0 , 
\qe 
where the unitary
matrix $B$ satisfies 
\eq 
\Gamma_M^\star = B\Gamma B^{-1} 
\qe
and in $11$ dimensions one can choose 
\eq
B = B^\star, \qquad B = B^\dagger .  
\qe 
In the representation chosen above for the Dirac matrices, one may take 
\eq
B=\Gamma_2 \Gamma_5 \Gamma_7 \Gamma_8 \Gamma_{10} .  
\qe
Let now
\eq
\eta = \alpha\otimes \ket{000} .
\qe
This satisfies
\eqr
i\Gamma_0\Gamma_1\Gamma_2 \Gamma_3\eta &=&\eta , \nonumber \\
\Gamma_{\bar{m}} \Gamma_{n} \eta &=& 0 \qquad (m\neq n)
\rqe
so it has $4$ components (and positive $4$-dimensional chirality). 
It is easy to see that $B\chi^\star$ has $(n_S,n_V,n_W)=(1,1,1)$ and
negative $4$-dimensional chirality. Thus the spinor 
\eq
\label{spinor}
\chi = \eta + B\eta^\star 
\qe
is a Majorana spinor which satisfies all the conditions. It has $4$ real
degrees of freedom, as expected for N=1 supersymmetry.

\sect{Fivebrane-Fivebrane Intersections}

According to supergravity calculations \cite{gau,tse} a pair of fivebranes
can meet along 
a $3$ space or a $1$-space. The intersection will appear as a
state in the spectrum of the quantum field theory on the ``vacuum''
fivebrane. Depending on the geometry of the intersection this state might 
appear as a membrane, a string or a particle. To have stable BPS states
these configurations must preserve some of the supersymmetries left intact
by the ``vacuum'' fivebrane. This section is devoted to checking whether
such supersymmetry preserving configurations can occur. 
%The calculations
%below do not actually determine the dimensionality of the intersection
%directly. 
The cases considered below treat the different possibilities for
the part of the intersection inside the $(0,1,2,3)$ space, that is, they
correspond to the different types of extended objects as viewed from the
four-dimensional worldvolume field theory. The actual dimensionality of an
intersection 
follows from solving the differential-geometric conditions which follow
from the calculations done here.

\subsection{Membrane Type Intersections}

Consider a membrane in MQCD represented by a fivebrane whose worldvolume
lies in $x^0, x^1, x^2$ and $M_3 \subset Y$ where $M_3$ is a three dimensional
manifold.  Thus the intersection is $R^2 \times (M_3\cap\Sigma)$, so from
the four-dimensional worldvolume field theory point of view it is a two
dimensional plane which divides physical space into two halves.  One
expects that $M_3 \cap \Sigma$ should be 
$1$-dimensional, since there are supergravity solutions of intersecting
fivebranes which intersect on three space
directions\cite{tow,gau,tse}. Further on an example is presented where this
is indeed the case. 

To preserve some supersymmetry there must be solutions of
\eq 
\chi =
\frac{1}{3!}\eps^{\alpha\beta\gamma} \Gamma_{0}\Gamma_{1}\Gamma_{2}
\Gamma_{ijk} 
\partial_{\alpha} X^i \partial_{\beta} X^j \partial_{\gamma} X^k \chi , 
\qe 
where now the $X$'s refer to the embedding of the second fivebrane. Writing
this in terms of holomorphic and anti-holomorphic coordinates as before,
and using the fact that
$\Gamma_{\bar{m}}\Gamma_{n}~\chi = 0$ if $m\neq n$ one finds
\eqr
\chi&=&\eps^{\alpha\beta\gamma}
\Gamma_{0}\Gamma_{1}\Gamma_{2} 
(\Gamma_{mnp}\partial_{\alpha}X^{m}\partial_{\beta}X^{n}\partial_{\gamma}X^{p}
\nonumber \\
&+& 6\Gamma_{mn\bar{p}}\partial_{\alpha}X^{m}\partial_{\beta}X^{n}
\partial_{\gamma}X^{\bar{p}} + h.c)\chi , 
\rqe 
Using \req{spinor} and matching terms with the same ``occupation numbers''
on both sides of this equation gives
\eq
B\chi^\star\eps_{\alpha\beta\gamma} =  \Gamma_{0}\Gamma_{1}\Gamma_{2}
\Gamma_{mnp}\partial_{\alpha}X^{m}\partial_{\beta}X^{n}
\partial_{\gamma}X^{p}\chi 
\qe
and
\eq
0=\Gamma_{0}\Gamma_{1}\Gamma_{2}\Gamma_m \Gamma_{n\bar{n}} 
\partial_{\alpha}X^{m}\partial_{\beta}X^{n}\partial_{\gamma}X^{\bar{n}}\chi
\qe 
(no sum over $m$ in the last equation). 

The first of these conditions can be rewritten as\footnote{$X_\star\Omega$
is the pullback of the holomorphic threeform to $M_3$.} 
\eq
\omega B\eta^\star = (X_\star\Omega) \Gamma_{0}\Gamma_{1}\Gamma_{2}
\Gamma_{SVW}\eta ,
\qe
where $\omega$ is the volume form on $M_3$, while $\Omega$ is 
the holomorphic $3$-form 
\eq
\Omega = \frac{dt}{t} \wedge dv \wedge dw .
\qe
Thus the pullback of the holomorphic three-form to $M_3$ must be equal to
the volume form on $M_3$.  

The second equation states that the
pullback of the K\"ahler form to $M_3$ must vanish: this follows from the
fact that
\eq
\Gamma_{m\bar{n}} \eta = J_{m\bar{n}} \eta .
\qe

Since $\Omega$ is equal to the volume form, one concludes that for a
BPS saturated membrane the tension is given by
\eq
\label{tens}
T = |\int_{M_3} \Omega| ,
\qe
which can be computed for any suitable brane configuration. It is a
universal formula for the membrane tension.

\subsection{String Type Intersections}

In this case the fivebrane has its worldvolume filling $(x^0, x^1)\times
M_4$, where $M_4$ is a submanifold in $Y$. The condition for supersymmetry
preservation reads in this case
\eq 
\label{string}
\chi =
\frac{1}{4!}\eps^{\alpha\beta\gamma\delta} \Gamma_{0}\Gamma_{1}
\Gamma_{ijkl} 
\partial_{\alpha} X^i \partial_{\beta} X^j \partial_{\gamma} X^k
\partial_{\delta} X^l \chi .
\qe 
Writing
this in terms of holomorphic and anti-holomorphic coordinates as before,
and using the fact that
$\Gamma_{\bar{m}}\Gamma_{n}~\chi = 0$ if $m\neq n$ one finds
\eqr
\eps^{\alpha\beta\gamma\delta}\chi&=&\Gamma_{0}\Gamma_{1}
(4\Gamma_{mnp\bar{q}}\partial_{\alpha}X^{m}\partial_{\beta}X^{n} 
\partial_{\gamma}X^{p} \partial_{\gamma}X^{\bar{q}} 
\nonumber \\
&+& 6\Gamma_{mn\bar{p}\bar{q}}\partial_{\alpha}X^{m}\partial_{\beta}X^{n}
\partial_{\gamma}X^{\bar{p}}\partial_{\delta} X^{\bar{q}} 
\nonumber \\
&+&4\Gamma_{m\bar{n}\bar{p}\bar{q}} \partial_{\alpha}X^{m}
\partial_{\beta}X^{\bar{n}} \partial_{\gamma}X^{\bar{p}}
\partial_{\delta}X^{\bar{q}} 
)\chi , 
\rqe 
Using \req{spinor} and matching terms with corresponding ``occupation
numbers'' on both sides of 
this equation gives
\eqr
\eta&=&{1\over 4} \eps^{\alpha\beta\gamma\delta} \Gamma_{0}\Gamma_{1} 
\Gamma_{mn\bar{p}\bar{q}}\partial_{\alpha}X^{m}\partial_{\beta}X^{n}
\partial_{\gamma}X^{\bar{p}}\partial_{\delta} X^{\bar{q}}\eta ,
\nonumber\\
0&=&\eps^{\alpha\beta\gamma\delta}\Gamma_{0}\Gamma_{1}
(\Gamma_{mnp\bar{q}}\partial_{\alpha}X^{m}\partial_{\beta}X^{n} 
\partial_{\gamma}X^{p} \partial_{\delta}X^{\bar{q}} \eta
\nonumber \\
&+&\Gamma_{m\bar{n}\bar{p}\bar{q}} \partial_{\alpha}X^{m}
\partial_{\beta}X^{\bar{n}} \partial_{\gamma}X^{\bar{p}}
\partial_{\delta}X^{\bar{q}} B\eta^\star) .
\rqe
Since 
\eq
\Gamma_{mn\bar{p}\bar{q}}\eta = (g_{n\bar{p}} g_{m\bar{q}} - g_{n\bar{q}}
g_{m\bar{p}}) \eta ,
\qe
the first of the above conditions implies
\eq
\omega = J\wedge J .
\qe
Since $\Gamma_0\Gamma_1\Gamma_{mnp\bar{q}}$ and
$\Gamma_0\Gamma_1\Gamma_{mnp\bar{q}}$ have opposite chiralities, the second
condition leads to $J=0$, so this solution is singular. It is
not clear whether this has any interesting interpretation.

\subsection{Particle Type Intersections}

The last case to consider is when the two fivebranes have a $0$-dimensional
intersection, which would correspond to a particle state in the worldvolume
field theory. It is easy to see that there are no BPS configurations of
this kind. The analog of \req{string} here would read
\eq
\eps_{\alpha\beta\gamma\delta\rho}\chi=\Gamma_{0}
(\Gamma_{mnp\bar{q}\bar{r}}\partial_{\alpha}X^{m}\partial_{\beta}X^{n} 
\partial_{\gamma}X^{p} \partial_{\delta}X^{\bar{q}}
\partial_{\rho}X^{\bar{r}} + h.c.) \chi .
\qe 
It is clear that the only solution to this is $\chi=0$, since there is no
way to match occupation numbers on both sides of this equation. This is
different from the case considered in \cite{fs}, where the curves
$\Sigma$ lead to an $N=2$ theory.

\section{An Example: $SU(n)$ Supersymmetric Yang-Mills}

This section presents an explicit example in $SU(n)$ supersymmetric
Yang-Mills theory. In the following section it will be argued that this
brane intersection corresponds to a domain wall in MQCD. The fivebrane
configuration here is quite nontrivial and is by itself a BPS state. The
$2+1$ dimensional field theory on its worldvolume may be interesting in its
own right.

\subsection{The Fivebrane Solution}
\label{example}

The vacuum fivebrane configurations \cite{wit2} are specified by Riemann
surfaces:
\eqr
t & = &v^n , \nonumber\\
vw & = &\zeta e^{i\frac{2m\pi}{n}},
\rqe
where $m$ labels the $n$ vacua of the $SU(n)$ gauge theory.  

Each vacuum fivebrane has a U(1) symmetry 
\eqr
&t&\longrightarrow e^{n i\delta } t ,\nonumber \\
&v&\longrightarrow e^{i\delta } v ,\nonumber \\
&w&\longrightarrow e^{- i\delta } w 
\rqe
and a $Z_2$ charge conjugation symmetry:
\eqr
\label{z2}
&v&\longrightarrow w ,\nonumber \\
&w&\longrightarrow v ,\nonumber \\
&t&\longrightarrow \zeta^n t^{-1} .
\rqe
The membrane solution constructed below will later be interpreted as a 
domain wall and so it will be required that the solution preserve these
symmetries  
since they are not spontaneously broken. This solution is constructed in
such a way that it interpolates between two different vacuum fivebranes
labeled by $m=0$ and $m=1$. 

The $3$-dimensional intersection of the two fivebranes ($M_3$) can be
parameterized by $\rho,\sigma,\theta$. To keep the U(1) intact the
fivebrane which plays the role 
of the wall must be then of the form 
\eqr 
v &=& f(\rho, \sigma) e^{i\theta} ,\nonumber \\ 
w &=& g(\rho, \sigma) e^{- i\theta} ,\nonumber \\ 
t &=& \exp(h(\rho,\sigma)) e^{i n \theta} .
\rqe
This form is prescribed uniquely by the $U(1)$ symmetry (up to
reparameterisation) where $h$ is real and $f,g$ are complex functions.  
Note that $\theta$ drops out at the intersection; it parameterises a
non-trivial cycle on $\Sigma$. Without loss of generality
one can assume that one vacuum is attained for $\sigma=0$ and the other
one for  $\sigma=1$.

%$(\sigma,\rho)=(0,\rho_0)$
%and the other one for $(\sigma,\rho)=(1,\rho_1)$. 

The functions $f, g, h$ can be determined from the BPS conditions. 
The K\"ahler form on $Y$ reads
\eq
J = \vol{v} + \vol{w} + \frac{1}{|t|^2} \vol{t} .
\qe
The pullback of this to $M_3$ is
\eq
J = \vol{f} + \vol{g} + i[d(|g|^2-|f|^2)-dh]\wedge d\theta.
\qe
To satisfy $J_{\rho\theta}=J_{\sigma\theta}=0$ one has to take
\eq
h = \frac{1}{2n} (|g|^2-|f|^2) + C
\qe
for some constant $C$. This leaves
\eq
J = \vol{f} + \vol{g}.
\qe
The functions $f,g$ must be chosen to make this vanish. 

The pullback of the holomorphic 3-form $\Omega$ reads
\eq
\Omega = in df\wedge dg\wedge d\theta - i d(fg)\wedge d\theta\wedge dh
\qe
Since the BPS conditions require that this be equal to the volume form (up
to a constant phase), one must ensure that this indeed has a constant
phase. 

The Ansatz considered in the following reads: 
\eqr
f(\rho,\sigma) &=& u(\sigma) e^{i\rho} , \nonumber \\
g(\rho,\sigma) &=& u(\sigma) e^{-i\rho} 
\rqe
for some complex function $u$. This Ansatz ensures that $J=0$, and 
$\Omega$ simplifies to 
\eq
\Omega = n (\partial_\sigma u^2) d\sigma\wedge d\rho\wedge d\theta .
\qe
For this to have a constant phase $u^2$ must be linear in $\sigma$:
\eq
u=\sqrt{a+b\sigma}
\qe
for some constants $a,b$ which can be fixed by imposing the boundary
conditions:
\eqr
\sigma=0 &\Rightarrow& wv = \zeta ,\nonumber \\
\sigma=1 &\Rightarrow& wv = \zeta e^{\frac{2\pi i}{ n}} .
\rqe
This leads to the final form of the curve
\eqr 
t &=& \zeta^{n/2} e^{i n \theta} ,\nonumber \\ 
v &=& \zeta^{1/2} \sqrt{1-\sigma (1-e^{\frac{2\pi i}{ n}})}
e^{i(\rho+\theta)} ,\nonumber \\ 
w &=& \zeta^{1/2} \sqrt{1-\sigma (1-e^{\frac{2\pi i}{ n}})}
e^{-i(\rho+\theta)}. 
\rqe
Here the constant $C$ has been fixed by requiring the $Z_2$ symmetry
\req{z2}.

It is easy to check that this curve satisfies the BPS condition which
equates $\Omega$ with the volume form. It follows directly that 
\eq
\Omega = n \sin(2\pi/n) d\rho\wedge d\sigma \wedge d\theta ,
\qe
while the determinant of the metric on $M_3$ is easily found to be
\eq
det(g) = |\partial_\sigma u^2|^2
\qe
and the BPS condition is satisfied. 

Finally observe that this configuration has actually two independent $U(1)$
symmetries, 
\eqr
\rho \longrightarrow \rho &+& \delta_1 ,\nonumber \\
\theta \longrightarrow \theta &+& \delta_2
\rqe
for real $\delta_1,\delta_2$, but only the second one is also a symmetry of
the ``vacuum'' fivebrane.

\subsection{Interpretation}

Domain walls in MQCD were discussed recently in \cite{wit2,matt,vol}. 
The realization explored there seems to be different than the one presented
here. The intersecting fivebrane configuration found in the previous
section suggests a picture where a domain wall is realized as a piece of
fivebrane crossing another fivebrane. 

The solution given by \req{example} describes two fivebranes which
intersect along a $3$-space: two of the dimensions are non-compact and
appear as a membrane in the $4$-dimensional quantum field theory of the
vacuum fivebrane worldvolume, while the remaining dimension is a cycle on
$\Sigma$. The fivebrane which intersects the ``vacuum'' fivebrane
interpolates between this vacuum at $\sigma=0$ (where the branes intersect)
and an adjacent vacuum at $\sigma=1$. The physical picture is that as the
plane $x^3=0$ is crossed the ``vacuum'' fivebrane changes the configuration
of the compact part from the $m=0$ vacuum (at $\sigma=0$) to the $m=1$
vacuum at $\sigma=1$ on the other side. Thus one may think of one
fivebrane patched from three pieces: the vacuum fivebrane with $m=0$, the
membrane and the vacuum fivebrane with $m=1$. This is probably the
connection with Witten's picture of a single fivebrane which interpolates
between the different MQCD vacua. 

When the branes are separated in the $x^9$ direction
there is of course no 
domain wall in the $3+1$ dimensional worldvolume field theory. It only
appears when the separation in the $x^9$ direction vanishes. It is possible
to infer that a marginal bound state of two fivebranes appears
when they intersect on a three-dimensional manifold.  The argument 
relates this system to a D4-D0 brane system at least for flat branes
\footnote{We would like
to thank Oren Bergman and Albion Lawrence for discussions on this
question.}. Consider a pair of M-theory
fivebranes with worldvolumes along $(0,1,2,3,4,10)$ and
$(0,1,2,6,7,10)$. By compactifying $x^{10}$ and T-dualising along
$(1,2,6,7)$ one ends up with a D-fourbrane with worldvolume along
$(0,3,4,6,7)$ and a D-zerobrane. This system is known to form a bound state
at threshold\cite{set}.  Thus a pair of fivebranes with
a $3$-dimensional intersection forming a bound state is consistent with the 
duality argument just presented. 

Having the explicit brane configuration makes it possible to calculate the
tension explicitly using this and formula \req{tens}. This yields
(reinstating $R$) 
\eq
T = 4 \pi^2 n R T_5 |\zeta|^2 |1-e^{2\pi i/n}|^2 .
\qe
It is the same as that derived in \cite{wit2}, up to a constant factor
which is not fixed by Witten's analysis. The present approach fixes the
normalisation unambiguously.  

It should be possible to understand this result from the point of view of 
the brane worldvolume
field theory. It is clear that one is dealing with a membrane in the
non-compact space where the field theory is defined, but one would like to
understand in physical terms why it is that the vacua on either side of
the membrane are different. 

One can actually see that the membrane couples in the worldvolume field
theory as a domain wall.  Consider the $3+1$ dimensional worldvolume theory on
the ``vacuum'' fivebrane. 
The fivebrane worldvolume theory is a six
dimensional field theory with 16 real 
supersymmetries, whose field content is the (2,0) tensor
multiplet. This contains $5$ real scalars, interpreted as fivebrane
transverse positions. In \cite{tow} it was argued that the 3-space intersection
of two fivebranes couples to a four-form gauge potential obtained by
dualising the 1-form field strength of a linear combination of the five
scalars $\phi$ in the 
worldvolume theory (the correct combination of scalars depends on the 
orientation of the intersection):
\eq
\star d\phi = F_5 = dA_4 .
\qe
The action for this system will contain terms like
\eq
\label{act6}
S = \int_{(0,1,2,3)\times \Sigma} \star F_5 \wedge F_5 + \mu_0
\int_{(0,1,2)\times \Sigma \cap M_3} A_4 .
\qe
The reduced theory on the $3+1$ dimensional non-compact worldvolume will
have in the action terms induced by \req{act6}. They read
\eq
S = \int_{(0,1,2,3)} \star F_4 \wedge F_4 + \mu \int_{(0,1,2)} A_3 .
\qe
where 
\eq
A_4 = A_3\wedge \xi .
\qe
Here $\xi$ is a cohomologically non-trivial one-form normalized so 
that
\eq
\int_{\Sigma} \xi \wedge\star\xi = 1
\qe
and 
\eq
\mu = \mu_0 \int_{\Sigma\cap M_3} \xi
\qe
is the effective charge. Since $\Sigma\cap M_3$ is the only non-trivial
cycle on $\Sigma$ and $\xi$ is the associated element of cohomology, this
does not vanish. The equation of motion reads
\eq
d\star F_4 = \mu \delta(x^3)
\qe
and requires that the scalar 
$\star F_4$ is locally constant, except at the wall where it undergoes a
jump by $\mu$. Thus its value on either side of the membrane is different,
which justifies regarding the membrane as a domain wall. 
This is analogous to the situation discussed in
\cite{polwit,bdgpt}, where an 8-brane acts as a domain wall in type IIA string
theory. 

One may therefore conclude that the ``wall'' fivebrane couples to the 
``vacuum'' fivebrane when the two branes are coincident. It would be
interesting to understand better the connection between this realisation
of an MQCD domain wall and the one studied in \cite{wit2,matt,vol}.

\sect{Twobrane-Fivebrane Intersections}

For completeness one may also consider the possible intersections of a
twobrane with  
the ``vacuum'' fivebrane. This could a priori lead to a BPS particle or 
string states in the worldvolume theory on the fivebrane 
It is very easy 
to see that there are no BPS configurations of this kind. To this end one
only needs to look at the analog of \req{memb}, which
would in the present case read
\eqr
\chi=\Gamma_{0} \Gamma_{1}(
\Gamma_{m}\partial_{\sigma}X^{m}+\Gamma_{\bar{m}}
\partial_{\sigma}X^{\bar{m}}) \chi
\rqe 
for the string case ($\sigma$ is the coordinate along the ``string''), and 
\eq
\eps_{\alpha\beta}\chi=\Gamma_{0}(
\Gamma_{mn}\partial_{\alpha}X^{m}\partial_{\beta}X^{n} 
+
\Gamma_{m\bar{n}}\partial_{\alpha}X^{m}\partial_{\beta}X^{\bar{n}}
+\Gamma_{\bar{m}\bar{n}}\partial_{\alpha}X^{\bar{m}} 
\partial_{\beta}X^{\bar{n}}
) \chi
\qe 
for the particle case. 

It is clear that the only solution to this is $\chi=0$, since there is no
way to match 'occupation numbers' on both sides of this equation.

\sect{Conclusions and Summary}

In this paper different possibilities of obtaining
extended objects in MQCD are explored.  The approach is to consider 
intersections of M-branes with the vacuum fivebrane
which preserve supersymmetry.  There are essentially
two possibilities, both of which involve a fivebrane intersecting
the vacuum fivebrane.  The intersections are such that
in the $(0,1,2,3)$ part of the vacuum worldvolume the intersection
looks like a membrane or a string.  The string solution is singular
and it is not clear whether there is anything interesting to be learned
from that case.  The case where the intersection looks like a membrane
is of greater interest and was studied in detail.   

A formula for the tension of the membrane was derived which is the integral
of the holomorphic three-form $\Omega$ 
over $M_3$ (the three dimensional part of the intersecting 
fivebrane's worldvolume embedded in the space $R^5\times S^1$).
The example of super-Yang-Mills was studied in detail and an
explicit 
example of a BPS saturated membrane was constructed and its tension was
calculated.
It was then argued that the membrane is in fact a domain wall in the
gauge theory living in the $0123$ directions of the vacuum fivebrane.
The tension agrees with the known result \cite{wit2} derived by
considering the difference in the vacuum density on either side
of a domain wall in supersymmetric Yang-Mills theory.  There the tension is
obtained by considering the difference in the
asymptotic values of the superpotential in $N=1$
supersymmetric Yang-Mills.  This is a means of calculating the
superpotential 
and may have wider applications.  

In \cite{vol} a vacuum fivebrane configuration which interpolates between
two vacua was constructed in the spirit of \cite{wit2}.  
The relation to the approach advocated in
this paper lies in our conjecture that the vacuum fivebrane splits
along the intersecting fivebrane so that on either side of the intersecting
brane there are two different vacua.  Thus the configuration considered
here is probably a singular limit of \cite{vol}.  A more careful study of the
relationship of the two approaches is clearly of interest.

%\newpage
\vspace{0.5cm}

\begin{center}
  {\bf Acknowledgments}
\end{center}

We would like to thank Oren Bergman and Albion Lawrence for helpful
discussions. We are also indebted to Andrew Strominger for very inspiring
questions and suggestions.

The work of M.S. was supported by a Fulbright Fellowship.

\newpage               

\newcommand{\bi}[1]{\bibitem{#1}}

\newcommand{\plb}[3]{{\em {Phys. Lett.}} {\bf B {#1}} (19{#2}) {#3}} 
\newcommand{\npb}[3]{{\em {Nucl. Phys.}} {\bf B {#1}} (19{#2}) {#3}} 
\newcommand{\prl}[3]{{\em Phys. Rev. Lett.} {\bf {#1}} (19{#2}) {#3}} 
\newcommand{\cmp}[3]{{\em {Comm. Math. Phys.}} {\bf {#1}} (19{#2}) {#3}} 
\newcommand{\mpla}[3]{{\em Mod. Phys. Lett.} {\bf A {#1}} (19{#2}) {#3}.}
\newcommand{\ijmpa}[3]{{\em Int. J. Mod. Phys.} {\bf A {#1}} (19{#2}) {#3}}

\end{document}